# Macroscopic uniform 2D moiré superlattices with controllable angles

Gregory Zaborski Jr.[1], Paulina E. Majchrzak[2], Samuel Lai[1], Amalya C. Johnson[1], Ashley P. Saunders[3], Ziyan Zhu[4], Yujun Deng[4,5], Donghui Lu[4,5], Makoto Hashimoto[4,5], Z-X Shen[2,4,5,6], Fang Liu[3*]

[1]Department of Materials Science and Engineering, Stanford University, Stanford, CA, 94305, USA

[2] Department of Applied Physics, Stanford University, Stanford, CA, USA

[3] Department of Chemistry, Stanford University, Stanford, CA, 94305, USA

[4] Stanford Institute for Materials and Energy Sciences, SLAC National Accelerator Laboratory, Menlo Park, CA, USA

[5] SLAC National Accelerator Laboratory, Menlo Park, California 94025, USA.

[6] Department of Physics, Stanford University, Stanford, CA, USA

## Abstract

Moiré superlattices, engineered through precise stacking of van der Waals (vdW) layers, hold immense promise for exploring strongly correlated[1–4] and topological phenomena[5,6]. However, these applications have been held back by the common preparation method: tear-and-stack[7] of Scotch tape exfoliated monolayers. It has low efficiency and reproducibility[8], along with challenges of twist angle inhomogeneity, interfacial contamination[9], micrometer sizes[8], and a tendency to untwist at elevated temperatures[10]. Here we report an effective strategy to construct highly consistent vdW moiré structures with high production throughput, near-unity yield, pristine interfaces, precisely controlled twist angles, and macroscopic scale (up to centimeters) with enhanced thermal stability. We further demonstrate the versatility across various vdW materials including transition metal dichalcogenides, graphene, and hBN. The expansive size and high quality of moiré structures enables high-resolution mapping of the reciprocal space back-folded lattices and moiré mini band structures with low energy electron diffraction (LEED) and angle-resolved photoemission spectroscopy (ARPES). This technique will have broad applications in both fundamental studies and mass production of twistronic devices.

## Main

Moiré superlattices arise from interference at the interface between two crystal lattice planes differing in lattice constants and/or alignment angles. With tunable band-filling and doping conditions, moiré superlattices became a versatile platform to study the collective behavior of electrons[11], excitons[12], solitons[13], and topological band structures.[6,14] At specific twist angles (i.e. magic angle) of a van der Waals (vdW) bilayer interface, these superlattices substantially reduce electron kinetic energy, allowing Coulomb interactions to predominate, fostering strong electron correlations, resulting in flat electronic bands near the Fermi level.[15,16] In addition to bilayers, recent experimental developments are exploring moiré systems in mixed dimensional systems, with more robust superconductivity and richer excitonic physics[16–19]. For example, Van Hove singularities at magic angles were demonstrated for twisted graphene/graphite structures.[20] Recent transport measurements on the graphene/graphite system illustrated formation of a single quasi-two-dimensional hybrid structure through a combination of gate-tunable moiré potential and graphite surface states,[21,22] in which the properties of a bulk crystal are shown to be tuned by a superlattice potential occurring at the interface.

Achieving the exotic correlated electronic phases in a moiré structure demands very precise control of the rotational alignment between atomically thin layers, requiring angular precisions of ± 0.1° or less. However, reproducible fabrication of uniform moiré systems, particularly across macroscopic areas, remains a formidable challenge. The limitations encompass a broad spectrum of factors, including the uniformity of twist angles and strains[23], contamination-induced interfacial disorders[8,9], lattice relaxation/reconstruction[24,25] and the tendency of untwisting at high processing temperatures[10]. Experimental realization of vdW twisted structures has been accomplished with the tear-and-stack method using monolayers from traditional scotch tape exfoliaitons.[7] The exfoliation techniques are inherently stochastic, producing samples ranging from a few micrometers to tens of micrometers in size, limiting the area of the twisted moiré interfaces.[26][27] Moreover, these techniques often expose the monolayer to polymer and/or substrate surface prior to stacking, inevitably resulting in interfacial contamination and bubbles that weakens the bilayer interaction[8]. As a result, the efficiency and reproducibility of current tear-and-stack assembly methods is notably low. Although developments were made to perform tear and stack in vacuum environment[28,29], a monolayer surface still needs to touch a Si substate before bringing it into contact with another layer, with a risk to introduce substrate contamination. On the other hand, advanced growth techniques, such as chemical vapor deposition (CVD) or physical epitaxy, achieved direct synthesis of 2D monolayers over wafer scale[30]. Although recent advances have been made on growing microscopic bilayer flakes with different twist angles,[31] the precise manipulation of large-area twisted structures at controlled angles, particularly near the magic angle, remains elusive. Due to the challenges above in scalability and mass production, high-quality moiré structures are currently "more like an art piece" than a scalable product.[8] To harness the full potential of moiré materials, optimizing fabrication processes to ensure reproducibility, stability, and applicability across macroscopic areas is crucial.

The best-quality twisted vdW interface must be constructed by adhering two perfect 2D crystal surfaces. In this study, we introduce a deterministic approach to construct vdW heterostructures via in-situ cleavage and twisting directly at the original pristine vdW interface of bulk crystals. This technique produces highly uniform moiré superlattices from a wide range of 2D materials with unprecedented uniformity and versatility, achieving high throughput and near-unity yield over centimeter scale macroscopic areas. This advancement permits the precise determination and macroscopic exploration of moiré physics with spectroscopic tools that were previously limited by their spatial resolution. Furthermore, the macroscopic dimension will significantly enhance the thermal stability of the small twist angle structures, rendering them more robust against untwisting under high-temperature annealing processes.

**Rapid production of macroscopic moiré superlattices**

As depicted in Fig. 1, our approach entails the utilization of a fresh and clean gold surface templated-stripped from a plasma-cleaned silicon substrate. The gold surface has strong adhesion with a variety of vdW monolayers, offering capabilities of disassembling van der Waals crystals layer-by-layer into macroscopic monolayers.[32] To start the exfoliation, this templated stripped gold layer, backed with thermal-release tape (TRT) and polyvinylpyrrolidone (PVP) (Fig 1a), is stamped onto a freshly cleaved vdW crystal (Fig. 1b-c). Lifting the gold tape exfoliates a monolayer from the vdW crystal (Fig. 1d), which is subsequently rotated to a target angle (Fig. 1e) and stacked back onto the same vdW crystal (Fig. 1f) or another freshly cleaved vdW crystal. The lift-twist-drop off process is done within a couple of seconds with minimum exposure to the environment, creating twisted vdW interface *in situ* at the pristine vdW interface between the freshly cleaved monolayer and the freshly cleaved crystal surfaces. This swift process effectively addresses the interfacial contamination challenges inherent in traditional methods of stacking monolayers, which encompass contamination from polymers, solvents, and substrates. After assembly, the TRT, PVP, and Au components are meticulously removed through a sequence of thermal treatment, water rinsing, and etching with KI/I$_2$ solution, respectively (Fig. 1g), ensuring a clean top surface in the final moiré structure. Benefiting from the perfect lattice alignment of exfoliated monolayer within the original vdW layered crystal, the target twist angle in homo twisted moiré structures can be controlled very precisely, either manually or through automated processes.

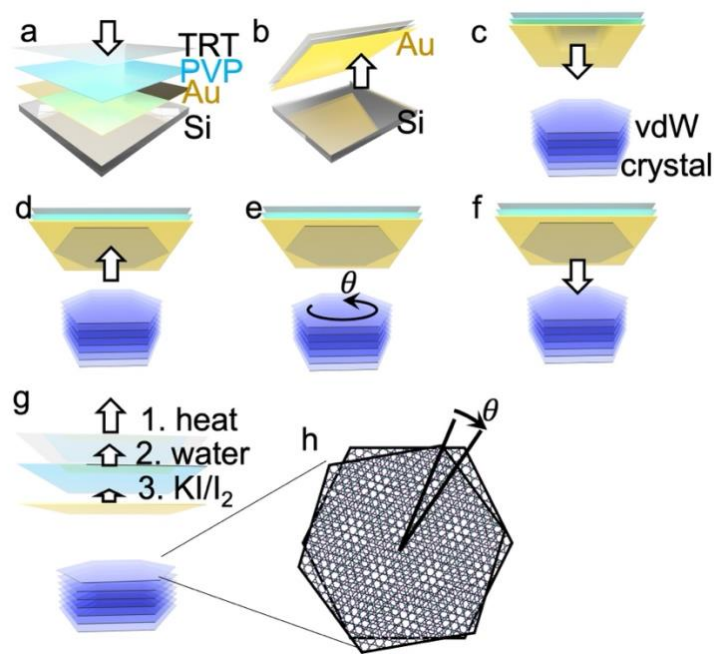

**Fig. 1 | Schematics for construction of macroscopic moiré structures**. **a** preparation of gold tape: depositing gold on an ultra-flat silicon wafer, spin-coating the surface with a PVP layer, and press thermal release tape (TRT) onto the stack. **b** template stripping the Au from Si. **c** pressing the ultra-flat gold onto the surface of a freshly cleaved bulk vdW crystal. **d** peeling off a monolayer. **e** rotating the monolayer at a fixed angle theta. **f** pressing the monolayer immediately back onto the bulk crystal. **g** removing the gold tape: removing the TRT with heat, dissolving PVP in water, and dissolving gold in a KI/I$_2$ etchant solution. **h** obtaining the moiré structures with macroscopic dimensions.

**Uniform, customizable 2D moiré superlattices over large length scales**

This *in situ* procedure is facile with near-unity exfoliation and stack efficiency. The lateral scale of twisted interface matches the bulk crystal's dimensions, representing a significant improvement compared to conventional techniques. Fig. 2 demonstrates an example of twisted MoS$_2$ monolayer/MoS$_2$ bulk system with cm-size dimension. The surface of the macroscopic moiré structure is flat with no bubbles, as directly created from pristine interface in the freshly cleaved vdW bulk crystal (additional images are shown in Fig. S1-2). We characterized the real-space moiré superlattice structures with piezoresponse force microscopy (PFM)[33] or torsional force microscopy (TFM)[34]. A noteworthy attribute is the exceptional precision in controlling twist angles over macroscopic areas, a capability derived from the rigid handling layer and the perfect lattice alignment in bulk vdW crystals. To illuminate the outstanding structural uniformity, we sampled four random locations on the surface. Despite the fact that these locations are over few millimeters apart, the moiré patterns are substantially consistent and uniform. From the periodicity of measured moiré structures, we can evaluate the corresponding twist angles, as labeled in Fig. 2. Remarkably, the twist angles vary by $< 0.1°$ over distance of millimeters. This unique nature allows for the creation of any arbitrary twisted systems with a high degree of versatility with superior accuracy over extended areas, a capability not available with conventional techniques.

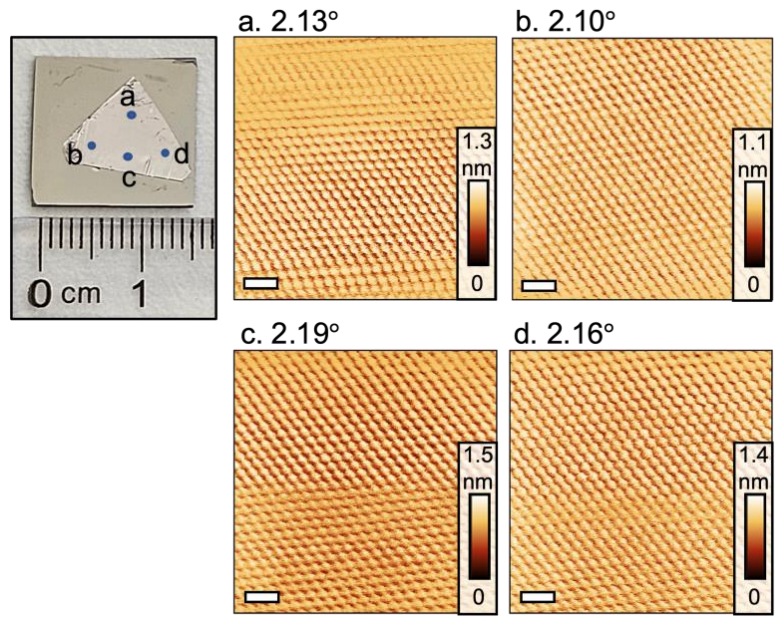

**Fig. 2 | Uniform moiré structures over macroscopic length scales. Left:** Picture showing cm size MoS₂/MoS₂ moiré superlattice on Si with four random locations scanned (a-d). **Right:** Lateral piezoresponse force microscopy (LPFM) images at each specific location. Scale bar: 25 nm. The moiré structures exhibit clear responses in both vertical and lateral PFM modes.

This technique allows effective creation of highly customizable twisted vdW interfaces across a diverse range of vdW materials. Example TFM/PFM images are presented in Fig. 3, demonstrating distinct moiré structures for different twisted vdW interfaces. Additional optical images and PFM/TFM images capturing more twist angles and/or over more extensive areas can be found in Fig. S3-S7. In addition to homo-twisted TMDC systems (such as MoS₂/MoS₂, WS₂/WS₂, MoSe₂/MoSe₂, WSe₂/WSe₂), combining a freshly exfoliated monolayer and another freshly cleaved vdW crystal leads to the production of twisted vdW heterostructures with similar uniformity over extended areas. A few examples of these moiré structures, including WSe₂/MoSe₂, WSe₂/WS₂, and WSe₂/MoS₂, are shown in Fig. 3 g-i. Moreover, we achieved large-area twisted graphene/graphite and hBN/hBN moiré structures, using the direct evaporation of gold[35,36] followed by a similar *in situ* exfoliation and stack process. These results highlight the method's efficacy and potential for exploring more complex vdW configurations in the future.

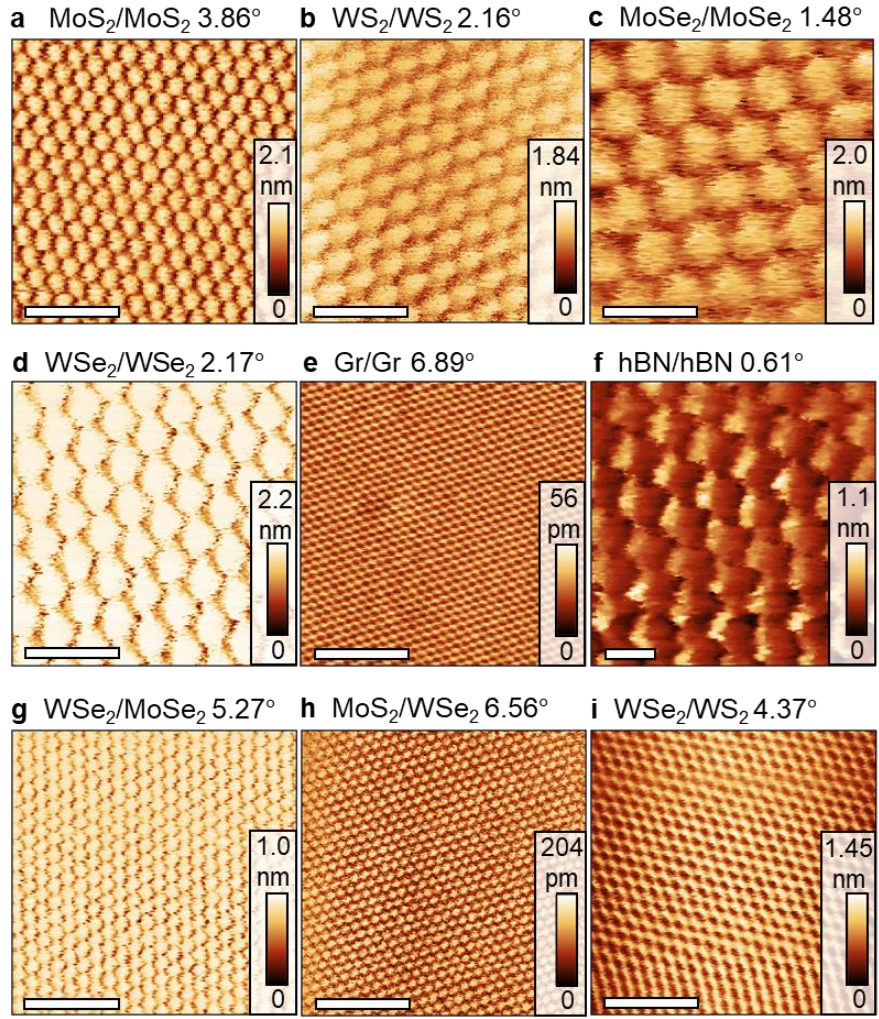

**Fig. 3 | TFM/PFM images of different monolayer-on-bulk moiré structures.** The angles correspond to the rotating angle θ in Fig. 1. The moiré superlattice measurements are conducted with imaging modes of: **a.-d.** and **g.-i.** lateral piezoresponse force microscopy; **e.** torsional force microscopy; **f.** vertical piezoresponse force microscopy. Scale bar: 25 nm.

**Periodic lattice distortion**

The uniformity of both the twist angle and moiré periodicity across extensive areas opens avenues for probing physics of moiré structures with a diverse array of tools that were previously limited by their spatial resolution. We characterized the reciprocal-space moiré superlattice structure with low energy electron diffraction (LEED), a conventional surface science technique with large electron beam spots up to 1 mm². The LEED spectra of a twisted $MoS_2/MoS_2$ is shown in Fig. 4a. The clear LEED diffraction image with a large electron beam spot showcases the consistency in twist angle and moiré periodicity across the macroscopic probe scale. The twist angles deduced from LEED are in good accordance with real-space lattice measurements through PFM. Apart from the Bragg peaks originating from the monolayer and bulk, the LEED diffraction patterns in Fig. 4a reveal distinct moiré superlattice diffraction peaks, notably in the vicinity of the monolayer diffraction pattern.

Moiré superlattices at the twisted vdW interface of 2D materials, particularly those with small twist angles, are frequently associated with picometer-scale periodic lattice distortions (PLDs). These phenomena involve both in-plane and out of plane lattice modulations, create alternating AB and BA stacked domains with soliton shear boundaries, resulting in reduction of local lattice symmetry and minimization of the total energy. PLDs are often accompanied by a charge density wave (CDW) or correlation related behaviors. Because of PLD, the reciprocal space lattice imaged by LEED is not simple superposition of two layers, but with additional superlattice (satellite) peaks in the moiré mini-Brillouin zone. The superlattice peaks reflects the long-periodic lattice restructuring forming a phase grating of the scattered electrons.

To infer a direct relationship between the intensity of superlattice peaks and the PLD amplitude, we simulated the torsional displacement field with a three transverse periodic lattice distortion model similar as that developed by Sung et al.[37] In the existing model, the monolayers at both sides of the vdW twist interface cooperate together with the same magnitude of PLD, optimizing the structural stability at the interface. In contrast, monolayer on bulk systems are not treated equally: because the monolayer is more flexible in adapting to induced strain than the bulk crystal, the monolayer predominantly adjusts the lattice to optimize commensurate domains and minimize energies. In contrast, the bulk crystal's lattice structure remains minimally impacted, due to the more rigid vdW bonding systems in the closed packed configuration of the bulk. Our simulation results are shown in Figure 4b and Fig. S8. When we restrict the reconstruction to the monolayer, our simulation result shows similar unbalanced intensities of moiré diffraction peaks around monolayer and bulk diffraction spots, in good accord with experimental LEED spectra.

A potential contribution to the superlattice signal in LEED is electron multiple scattering between monolayer and bulk crystal. To evaluate the origin of the superlattice signal, we analyzed the relative intensity of the superlattice peaks as a function of electron energy. In cases of pure multiple diffraction, the superlattice diffraction peaks will exhibit additional periodic intensity extinctions as a function of energy different than the main Bragg peaks, due to the in-phase and out-of-phase scattering condition.[38,39] As depicted in Fig. S9, although both superlattice peaks and main Bragg peaks are dependent on the electron beam energies, their relative intensities remain the same. This indicates that PLD is likely the major contribution to the superlattice structure, rather than mere multiple scattering between monolayer and bulk.

**Thermal stabilities**

At elevated temperatures, the small magic-angle twisted van der Waals (vdW) interface exhibits a tendency to revert from an incommensurate state to a zero-degree commensurate structure, the most thermodynamically stable configuration[10]. The thermally-activated untwisting results in instability of the moiré pattern and twist angles during sample fabrications. This presents a major challenge for the reproducibility in the experimental investigations of moiré physics and implementation of twistronic devices - the actual final twist angle often deviates from what was intended in the initial fabrication design. The untwisting process often navigates through multiple potential energy barriers, encountering several locally stable energy states[40] from the varying degrees of commensurability among atomic stackings across the twisted interface. Both the energy barrier and the number of locally stable energy states scale with the flake sizes.

As a result, the macroscopic twisted moiré structures will not untwist as easily as to the traditional microscopic moiré structures. Once lateral dimensions reach millimeter scale, the activation energy for untwisting becomes significantly higher than that of μm sized flakes, leading to a substantially improved thermal stability. To showcase thermodynamic rotational stability of the macroscopic twist vdW system, we heated a twisted $MoS_2$ sample to elevated temperature (500 °C) for an extended period of time >1h. The LEED patterns before and after the anneal process, as shown in Fig. S10, demonstrates identical twist angles before and after the high temperature treatment. The macroscopic moiré structures offer a particular route in achieving small-angle twisted moiré structures with substantial thermal stability, a great advantage for improving the reproducibility in future large-scale processing and device manufacturing.

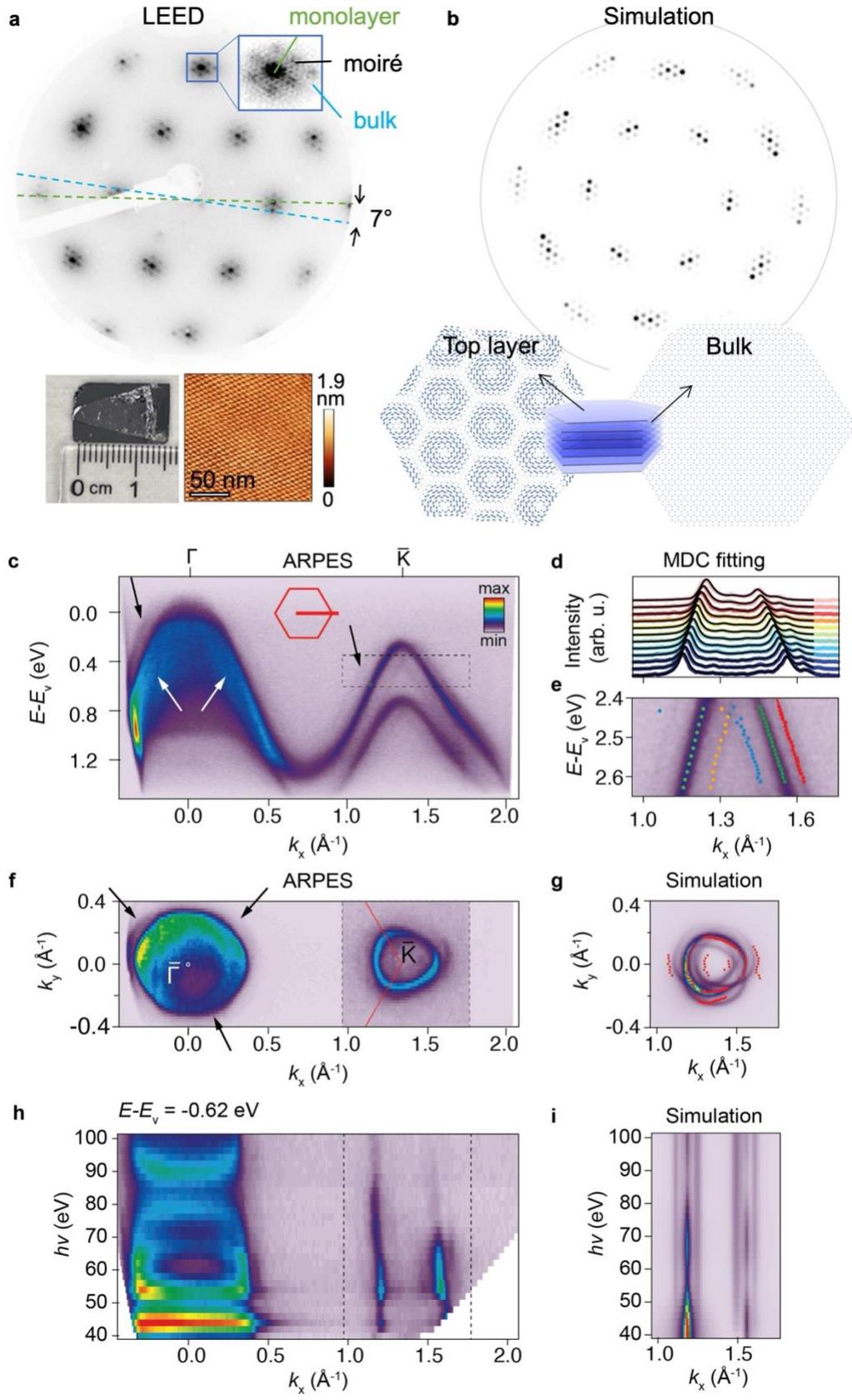

**a** LEED    monolayer    moiré    bulk    7°

**b** Simulation    Top layer    Bulk

1.9 nm    0 cm    1    50 nm    0

**c** ARPES    Γ    K̄    max    min    E−E_v (eV)    k_x (Å⁻¹)

**d** MDC fitting    Intensity (arb. u.)

**e** E−E_v (eV)    k_x (Å⁻¹)

**f** ARPES    k_y (Å⁻¹)    Γ°    K̄    k_x (Å⁻¹)

**g** Simulation    k_y (Å⁻¹)    k_x (Å⁻¹)

**h** E−E_v = −0.62 eV    hν (eV)    k_x (Å⁻¹)

**i** Simulation    hν (eV)    k_x (Å⁻¹)

**Fig. 4 | Reciprocal space characterization of moiré superlattices a.** Low-energy electron diffraction (LEED) of 7-degree twisted monolayer $MoS_2$/bulk $MoS_2$ moiré superlattice, collected with 142.3 eV electron energy. Bottom: Optical and LPFM images of the same sample. **b.** Simulated moiré LEED pattern constructed with predominant PLD on the top monolayer while the bulk crystal remains intact. Bottom: displacement field of the top layer and the bulk. **c.** High resolution angle-resolved photoemission (ARPES) spectrum along Γ-K high-symmetry line, taken at photon energy of 60 eV. Arrows point to the replica bands around the Γ- and K-points. **d.** Momentum-distribution curves (MDCs, markers) extracted from the region demarcated by the dashed box in c). Solid black lines are the fit results to the model of four Lorentzian peaks on a polynomial background. **e.** Band dispersion within the box in c). Colored markers indicate the fitted positions of the main and replica bands around the K-valley. **f.** Constant energy contour at the energy of $E$-$E_v = 0.62$ eV. Arrows indicate the replicas. The contrast has been enhanced around the K-valley. **g.** Spectral function simulation of the photoemission intensity in the cut corresponding to e). Two three-fold symmetric sets of replica bands are introduced, their intensity distribution cannot be predicted by Umklapp scattering. **h.** Experimental photon energy-dependence of the Γ-K cut at $E$-$E_v = 0.62$ eV. Both main and replica bands show nearly energy-independent narrow linewidths of comparable magnitudes, indicating a low degree of rotational disorder of the top layer. **i.** Simulated evolution of main and replica band intensity around K-valley with photon energy.

## Backfolded band structures

The introduction of a slowly varying periodic moiré potential leads to the formation of mini-Brillouin zones (BZs) with back-folded replicas of the electronic band structure. These replicas have been proposed to play a key role in the appearance of strongly correlated phases in semiconducting moiré materials by undergoing gap openings and flat band formation under the displacement field[1,41–43]. We directly observe the backfolded replicas in the moiré mini-BZ with angle resolved photoemission spectroscopy (ARPES). Fig. 4c displays the measured dispersion along the Γ-K high-symmetry line in a 4.5° $MoS_2$/$MoS_2$ twisted monolayer/bulk structure. Our high-quality twisted vdW interface over a large area enables us to discern faint satellite bands around both the Γ- and K-valleys, as highlighted by arrows. In contrast, previous ARPES studies of twisted TMDC homobilayers only reported mini-bands around the Γ center.[44,45] The difficulty of observing replica bands at K valley in the previous measurements was explained by the poor wavefunction overlap, due to the predominantly in-plane character of the Mo $d$-orbitals comprising these states[44,46]. There is a mounting body of evidence to suggest that these satellite bands arise due to the atomic reconstruction of the constituent layers rather than the interlayer interaction[47–49]. As demonstrated with the LEED measurements, the rigid underlying bulk crystal in our heterostructure is causing periodic atomic rearrangement of the top monolayer, creating distinct signature of K point replicas. While in our system it would be difficult to isolate these salient features in energy due to the dispersive nature of the underlying bulk, it is nevertheless a useful model system to study the triangular Hubbard Hamiltonian in the framework of a "quantum simulator"[51]. On the other hand, the satellite bands around the Γ-point are attributed to the interlayer interaction, which is expected to be enhanced by the high degree of cleanliness and uniformity of the interface in our sample[50]. Due to the broad manifold of states and significant out-of-plane dispersion at the Γ-point from the bulk MoS₂, a detailed analysis of the minibands at the Γ-point is challenging. Consequently, our analysis focuses on the moiré bands around the K-valley.

We track the dispersion at the K-valley through fitting of the momentum-distribution curves (MDC), marked on the spectrum in Fig. 4d. As revealed by the fit peak positions plotted in Figure 4e, the two satellite bands are symmetric about the main band, suggesting they are backfolded rather than originating from two adjacent domains. Both main and replica bands show nearly energy-independent linewidths of comparable magnitudes, 0.074 and 0.094 deg, respectively. The linewidth increase can be taken as a measure of rotational disorder of the top layer, indicating the upper limit of variation of ~0.05° across the beam spot of (30×5) μm², consistent with our PFM measurements.

To elucidate the origin of the replicas, we examine the constant energy contour (CEC) at the energy $E-E_v = 0.56$ eV. We clearly observe approximately threefold symmetric replica bands around the Γ- and K-valleys. The most prominent features of the CEC can be reproduced through spectral function simulations. As depicted in Fig. 4f and Fig. S11, the intensity distribution can be fully recreated by considering two sets of mini-bands displaced by the moiré reciprocal lattice vector for a twist angle θ = 4.5 degrees. Both sets exhibit C3 symmetry due to mirror-symmetry breaking induced by the in-plane distortions of the top monolayer[52]. One set follows the orientation of the moiré reciprocal vectors, while the other is rotated by 180 degrees. This can be explained by considering the electrons from the first and second topmost bulk layers, respectively, due to the 180-degree rotation between adjacent layers in a 2*H* crystal structure.

In general, the replica bands could arise from a combination of two effects: final-state umklapp scattering of the photoelectrons upon existing the surface,[53] which is independent of θ, and modification of initial Bloch states by the superlattice potential, which is more pronounced at small twist angles[53,54]. We perform matrix element analysis to evaluate their contributions qualitatively, using a layer and orbital interference model developed for bilayer MoS₂[55]. The initial-state effect is highly sensitive to the kinetic energy of the photoelectrons and orbital character of the emitter atoms.[56] The intensity modulation across the trigonally warped main band contour results from intra- and interlayer interference effects, affecting the local orbital make-up of the states around the K-valley[55,57,58]. In the Umklapp scattering scenario, the satellite bands should exhibit the same intensity distribution (including its relative orientation) as the original band[53]. However, as shown in the Fig. 4f-g and Fig. S11, our data contradict this pattern. Instead, we find that the outer replica features are best described by rotating the matrix element for the two sets of backfolded bands by 90 and 60 degrees, respectively. This observation underscores the significant influence of the superlattice potential, aligning with the small twist angle θ=4.5° which is within the range where correlated phenomena in TMDCs have been previously observed.[1,59]

We further evaluate the evolution of the band with varying photon energies hv. Fig. 4h demonstrates experimental photon energy-dependence of the Γ-K cut at $E-E_v = 0.62$ eV. Examples of full spectra at selected photon energies are presented in the Fig. S12. Comparing the experimental data to the simulation in Fig. 4i, our simulation qualitatively reproduces the intensity modulation of the main bands, capturing the intensity increase in the energy range 50-70 eV and suppression of the right branch at higher *hv*. This simulation also assumes that the minibands follow the same *hv*-dependence as the main bands. In comparison, the experimental moiré minibands exhibit enhancement of the rightmost replica around 60-75 eV, while the inner replicas retain approximately constant intensity across the entire investigated photon energy range. This result suggests that the observed distribution may emerge from a more complex moiré-induced effect than a simple final-state diffraction model.

**Conclusion**

In summary, we introduce a rapid and deterministic technique for engineering moiré superlattices of twisted vdW systems on a macroscopic scale with exceptional uniformity over mm to cm areas, which is only limited by size of the vdW crystal. It is versatile across a variety of vdW materials, making it adaptable for a broad range of twistronic applications. Moreover, the expansive size of moiré structures enhances the thermal stability of the twisted vdW structures. We also identified the back folded replica at both $\Gamma$ and K point directly in ARPES, resulting from the periodically modulating lattice deformation and interlayer interaction. The uniform structure will offer a promising pathway for leveraging more macroscopic static or dynamic scattering or spectroscopy techniques for twisted moiré structures, unlocking new fronters for studying hitherto inaccessible quantum phenomena in 2D moiré systems.

**Methods**

Sample preparation

Large area (mm-cm) moiré superlattices are prepared through an *in-situ* exfoliation and stacking process by leveraging a modified gold tape exfoliation method and subsequently restacking the freshly exfoliated monolayer within seconds back onto the bulk 2D material. Specifically, the gold stamps were fabricated by depositing 100-nm-thick gold onto a 4" silicon wafer (NOVA Electronic Materials LLC) using an automated high vacuum electron beam evaporator (Kurt J. Lesker LAB18). Subsequently a layer of polyvinylpyrrolidone (PVP) mixture (Millipore Sigma, molecular weight of 40,000 10 wt % in a solution of ethanol and acetonitrile wt 1/1) was spin coated on top of the deposited Au film for 2 minutes (400 rpm/s, 400 rpm). Then a piece of thermal release tape (TRT, Semiconductor Equipment Corp. RA95LS) was affixed to the prepared PVP/Au/silicon structure, for structural integrity and functioning as the release layer during the *in-situ* assembly process. The TRT/PVP/Au surface is stripped off the Si substrate, and subsequently pushed against a freshly cleaved bulk vdW crystal (HQ graphene). The gold surface peels off a monolayer from the bulk crystal, and is re-stacked back onto the freshly cleaved bulk surface within a couple of seconds to minimize the contamination. For preparation of graphene and hBN moiré structures, we evaporate Au directly on to a freshly cleaved graphite and hBN crystals with deposition rate of <0.5 Å /s, followed by spin coating with PVP. Exfoliations are performed with TRT to peel off PVP/Au/monolayer together, followed by swift restacking them on the freshly cleaved crystal.  The TRT layer is removed by heating in ~ 100 °C. The PVP layer is removed in deionized water rinses. Au layer is removed in KI/$I_2$ etchant solution [10 g of KI (99.9%, J.T. Baker) and 2.5 g of $I_2$ (99.99%, Spectrum) in 100 ml of deionized water] for 4 minutes.

Piezoresponse force microscopy (PFM) / Torsional force microscopy (TFM)

PFM measurements were performed using a Bruker Dimension Icon with Nanoscope V Controller in torsional and vertical deflection modes under room temperature. The drive amplitude was generally set to 1V with resonance frequencies of ~300kHz for vertical and ~700kHz for torsional PFM. Oxford Instruments ASYELEC-01-R2 Ti/Ir coated silicon probes of a force constant of 2.8 N/m were used. The atomic force microscopy micrograph data was mean plane subtracted and aligned using first degree polynomial. Moiré wavelengths measurements were extracted using fast Fourier transform using Gwyddion. The resultant wavelength was utilized to determine moiré twist angle. For TFM, we use the same probe and drive amplitude of 1V with a tip velocity of ~0.7 um/s and resonance frequency of ~700kHz.

Low Energy Electron Diffraction

The low-energy electron diffraction (LEED) is collected under room temperature with BDL800IR spectrometer, OCI Vacuum Microengineering, Inc, inside a custom build ultra-high vacuum chamber with a base pressure ~ $8 \times 10^{-10}$ Torr.

To calculate the reconstruction field of the monolayer to fit the LEED data, we use the equation for the torsional periodic lattice displacement (PLD) field defined in Sung et. al.[37],

$$\Delta_n = A_n \sum_{i=1}^{3} \widehat{A}_i \sin(n\mathbf{q_i} \cdot \mathbf{r_0} + \phi_i) ; \quad \widehat{A}_i \perp \mathbf{q_i}$$

where $\mathbf{r_0}$ are the undistorted atom positions, $\mathbf{q_i}$ is the PLD wave vector, and $\widehat{A}_i$ is the unit vector describing the transversity of the PLD. Three $\mathbf{q}$'s are 120° apart with a magnitude set by the twist angle ($|\mathbf{q}| \approx b\theta$, $b$ is the reciprocal lattice constant) to accommodate the symmetry of the moire pattern. The phase, $\phi_i$, shifts or alters the field, and is set to 0 in our simulations. The reconstructed lattice positions are given by $\mathbf{r} = \mathbf{r_0} + \mathbf{\Delta_n}$. We describe the system with a single torsional PLD ($\mathbf{\Delta_1}$) for a honeycomb lattice and ignore higher order harmonics. The amplitude of the field $A_1$ that generates the best fit is ~ 0.15 pm. To model reconstruction in a bilayer system, we apply equal and opposite PLD fields to two twisted monolayer lattices. To model reconstruction in monolayer on bulk system, we apply the PLD field only to monolayer lattice, and keep the bulk unreconstructed. Once the new atomic positions are calculated, diffraction is simulated by taking a 2D Fast Fourier Transform of the atomic positions.

Angled Resolved Photoemission Spectroscopy

High resolution ARPES measurements were performed at BL5-2 of Stanford Synchrotron Radiation Lightsource (SSRL), SLAC Linear Accelerator Laboratory, using horizontal linear light polarization equipped with a DA30-L electron analyzer. The angular resolution was set to 0.1° and the total energy resolution was set to 20 meV or better. The beam spot size was larger than 0.01 x 0.04 mm$^2$. Constant energy contours were collected at photon energy of 60 eV, while the photon energy scan between 40 and 100 eV. All the measurements were performed in ultra-high vacuum with a base pressure lower than $5 \times 10^{-11}$ Torr at the base temperature of ~7K. The materials are bonded to a doped Si substrate using silver epoxy EPO-TEK H21D. Prior to measurement, samples were annealed at 200 C for 1 hour.

ARPES simulation

The simulated photoemission intensity, $I(\mathbf{k}, \omega) = |M(\mathbf{k}, \omega)|^2 A(\mathbf{k}, \omega) f_{FD}(T, \omega)$, was obtained using the following form of the spectral function:

$$A(\mathbf{k}, \omega) = \frac{I_0}{\pi} \frac{\Sigma''(\mathbf{k}, \omega)}{[\hbar\omega - \varepsilon_{\mathbf{k}, \nu} - \Sigma'(\mathbf{k}, \omega)]^2 + \Sigma''(\mathbf{k}, \omega)^2}$$

where $f_{FD}(T, \omega)$ is the Fermi-Dirac distribution at electronic temperature $T$. The energy dispersion $\varepsilon_{\boldsymbol{k},\nu}$ and matrix element $|M(\boldsymbol{k}, \omega)|^2$ were calculated using the $k.p$ model[55] for a bilayer MoS$_2$. The parameters in the model were adapted with no adjustments. Once computed, the spectral function matrix was rotated and translated according to the operations described in the text, and the distributions for main and replica states were added together. The intensity, $I_0$, of the replica bands for first and second layer was set to 0.005 and 0.001 of the main band intensity, respectively. For simplicity, the imaginary part of self-energy, $\Sigma''(\boldsymbol{k}, \omega)$, was set to be the same for both moiré and main bands, while the real part, $\Sigma'(\boldsymbol{k}, \omega)$, was set to 0.

**Acknowledgements**
The material development is supported by the Defense Advanced Research Projects Agency (DARPA) under Agreement No. HR00112390108. G.Z.Jr. acknowledges support from Pat Tillman Foundation, and Diversifying Academia, Recruiting Excellence (DARE) fellowship through the Office of the Vice Provost for Graduate Education. P.E.M. acknowledges support from Stanford Energy Postdoctoral Fellowship through contributions from the Dai and Li Family Stanford Sustainability Postdoctoral Fellow Program Fund, Precourt Institute for Energy, Bits & Watts Initiative, StorageX Initiative, and TomKat Center for Sustainable Energy. S.L. acknowledges support from the Agency for Science, Technology and Research, Singapore. P.E.M., M.H., and Z.X.S., acknowledge the support of the U.S. Department of Energy, Office of Science, Office of Basic Energy Sciences, Division of Material Sciences and Engineering, under contract DE-AC02-76SF00515. Use of the Stanford Synchrotron Radiation Lightsource, SLAC National Accelerator Laboratory, is supported by the U.S. Department of Energy, Office of Science, Office of Basic Energy Sciences under Contract No. DE-AC02-76SF00515. Use of the Stanford Nano Shared Facilities (SNSF), is supported by the National Science Foundation under award ECCS-2026822.

The authors thank Dr. Anil Rajapitamahuni and Dr. Elio Vescovo at Brookhaven National lab; Dr. Jonathan Sobota at SLAC; and Dr. Markus Scholz at DESY Germany, for their generous support in the design and early development of the ARPES measurements.

**Author contributions**
F.L. conceived this project. G.Z.Jr. fabricated all the samples and performed all the TFM/PFM measurements and analysis. P.E.M., F.L., M.H., and G.Z.Jr. performed ARPES measurements with the support from Y.D., and D.L.. P.E.M. performed simulation and interpretation of the ARPES data with the support from Z.Z.. Z.X.S. supervised the ARPES experiments and the data interpretation. F.L., S.L. and A.P.S. performed the LEED measurement and data analysis. A.C.J. performed the simulation of PLD and diffraction images. G.Z.Jr., P.E.M. and F.L. wrote the paper with input from all authors.

**References**
1.   Wang, L. *et al.* Correlated electronic phases in twisted bilayer transition metal dichalcogenides. *Nat Mater* **19**, 861–866 (2020).
2.   Turunen, M. *et al.* Quantum photonics with layered 2D materials. *Nature Reviews Physics* vol. 4 219–236 Preprint at https://doi.org/10.1038/s42254-021-00408-0 (2022).

3.      Cao, Y. *et al.* Unconventional superconductivity in magic-angle graphene superlattices. *Nature* **556**, 43–50 (2018).

4.      Sharpe, A. L. *et al. Emergent Ferromagnetism near Three-Quarters Filling in Twisted Bilayer Graphene*. https://www.science.org.

5.      Liu, X. & Hersam, M. C. 2D materials for quantum information science. *Nature Reviews Materials* vol. 4 669–684 Preprint at https://doi.org/10.1038/s41578-019-0136-x (2019).

6.      Xia, L.-Q. *et al.* Helical trilayer graphene: a moir\'e platform for strongly-interacting topological bands. (2023).

7.      Kim, K. *et al.* Van der Waals Heterostructures with High Accuracy Rotational Alignment. *Nano Lett* **16**, 1989–1995 (2016).

8.      Lau, C. N., Bockrath, M. W., Mak, K. F. & Zhang, F. Reproducibility in the fabrication and physics of moiré materials. *Nature* vol. 602 41–50 Preprint at https://doi.org/10.1038/s41586-021-04173-z (2022).

9.      Uri, A. *et al.* Mapping the twist-angle disorder and Landau levels in magic-angle graphene. *Nature* **581**, 47–52 (2020).

10.     Wang, D. *et al.* Thermally Induced Graphene Rotation on Hexagonal Boron Nitride. *Phys Rev Lett* **116**, (2016).

11.     Park, H. *et al.* Observation of fractionally quantized anomalous Hall effect. *Nature* **622**, 74–79 (2023).

12.     Wilson, N. P., Yao, W., Shan, J. & Xu, X. Excitons and emergent quantum phenomena in stacked 2D semiconductors. *Nature* vol. 599 383–392 Preprint at https://doi.org/10.1038/s41586-021-03979-1 (2021).

13.     Ni, G. X. *et al.* Soliton superlattices in twisted hexagonal boron nitride. *Nat Commun* **10**, (2019).

14.     Chaves, A. *et al.* Bandgap engineering of two-dimensional semiconductor materials. *npj 2D Materials and Applications* vol. 4 Preprint at https://doi.org/10.1038/s41699-020-00162-4 (2020).

15.     Suárez Morell, E., Correa, J. D., Vargas, P., Pacheco, M. & Barticevic, Z. Flat bands in slightly twisted bilayer graphene: Tight-binding calculations. *Phys Rev B Condens Matter Mater Phys* **82**, (2010).

16.     Bistritzer, R. & MacDonald, A. H. Moiré bands in twisted double-layer graphene. *Proc Natl Acad Sci U S A* **108**, 12233–12237 (2011).

17.     Halbertal, D. *et al.* Moiré metrology of energy landscapes in van der Waals heterostructures. *Nat Commun* **12**, (2021).

18.     Yu, L. *et al.* Observation of quadrupolar and dipolar excitons in a semiconductor heterotrilayer. *Nat Mater* **22**, 1485–1491 (2023).

19.     Li, W. *et al.* Quadrupolar–dipolar excitonic transition in a tunnel-coupled van der Waals heterotrilayer. *Nat Mater* **22**, 1478–1484 (2023).

20.     Li, G. *et al.* Observation of Van Hove singularities in twisted graphene layers. *Nat Phys* **6**, 109–113 (2010).

21.     Mullan, C. *et al.* Mixing of moiré-surface and bulk states in graphite. *Nature* **620**, 756–761 (2023).

22.     Waters, D. *et al.* Mixed-dimensional moiré systems of twisted graphitic thin films. *Nature* **620**, 750–755 (2023).

23.     Bai, Y. *et al.* Excitons in strain-induced one-dimensional moiré potentials at transition metal dichalcogenide heterojunctions. *Nat Mater* **19**, 1068–1073 (2020).

24. Yoo, H. *et al.* Atomic and electronic reconstruction at the van der Waals interface in twisted bilayer graphene. *Nat Mater* **18**, 448–453 (2019).

25. Weston, A. *et al.* Atomic reconstruction in twisted bilayers of transition metal dichalcogenides. *Nat Nanotechnol* **15**, 592–597 (2020).

26. Novoselov, K. S. *et al. Electric Field Effect in Atomically Thin Carbon Films. Phys. Rev. Lett* vol. 404 www.arXiv.org/quant-ph/ (2000).

27. de Jong, T. A. *et al.* Imaging moiré deformation and dynamics in twisted bilayer graphene. *Nat Commun* **13**, (2022).

28. Kang, K. *et al.* Layer-by-layer assembly of two-dimensional materials into wafer-scale heterostructures. *Nature* **550**, 229–233 (2017).

29. Mannix, A. J. *et al.* Robotic four-dimensional pixel assembly of van der Waals solids. *Nat Nanotechnol* **17**, 361–366 (2022).

30. Cai, Z., Liu, B., Zou, X. & Cheng, H. M. Chemical Vapor Deposition Growth and Applications of Two-Dimensional Materials and Their Heterostructures. *Chemical Reviews* vol. 118 6091–6133 Preprint at https://doi.org/10.1021/acs.chemrev.7b00536 (2018).

31. Sun, L. *et al.* Hetero-site nucleation for growing twisted bilayer graphene with a wide range of twist angles. *Nat Commun* **12**, (2021).

32. Liu, F. *et al. Disassembling 2D van Der Waals Crystals into Macroscopic Monolayers and Reassembling into Artificial Lattices.* https://www.science.org.

33. McGilly, L. J. *et al.* Visualization of moiré superlattices. *Nat Nanotechnol* **15**, 580–584 (2020).

34. Pendharkar, M. *et al.* Torsional Force Microscopy of Van der Waals Moir\'es and Atomic Lattices. (2023) doi:10.1073/pnas.2314083121.

35. Moon, J.-Y. *et al. Layer-Engineered Large-Area Exfoliation of Graphene. Sci. Adv* vol. 6 https://www.science.org (2020).

36. McKeown-Green, A. S. *et al.* Millimeter-Scale Exfoliation of hBN with Tunable Flake Thickness for Scalable Encapsulation. *ACS Appl Nano Mater* **7**, 6574–6582 (2024).

37. Sung, S. H. *et al.* Torsional periodic lattice distortions and diffraction of twisted 2D materials. *Nat Commun* **13**, (2022).

38. Tiefenbacher, S., Pettenkofer, C. & Jaegermann, W. *Moiré Pattern in LEED Obtained by van Der Waals Epitaxy of Lattice Mismatched WS 2 /MoTe 2 (0001) Heterointerfaces. Surface Science* vol. 450 www.elsevier.nl/locate/susc (2000).

39. Hoegen M. Horn-von Hoegen, Al-Falou A., Pietsch H., Müller B.H. & Henzler M. *Formation of Interfacial Dislocation Network in Surfactant Mediated Growth of Ge on Si(111) Investigated by SPA-LEED. Surface Science* vol. 298 (1993).

40. Bagchi, S., Johnson, H. T. & Chew, H. B. Rotational stability of twisted bilayer graphene. *Phys Rev B* **101**, (2020).

41. Naik, M. H. & Jain, M. Ultraflatbands and Shear Solitons in Moiré Patterns of Twisted Bilayer Transition Metal Dichalcogenides. *Phys Rev Lett* **121**, (2018).

42. Wu, F., Lovorn, T., Tutuc, E., Martin, I. & Macdonald, A. H. Topological Insulators in Twisted Transition Metal Dichalcogenide Homobilayers. *Phys Rev Lett* **122**, (2019).

43. Huang, M. *et al.* Giant nonlinear Hall effect in twisted bilayer WSe2. *Natl Sci Rev* **10**, (2023).

44. Pei, D. *et al.* Observation of Γ -Valley Moiré Bands and Emergent Hexagonal Lattice in Twisted Transition Metal Dichalcogenides. *Phys Rev X* **12**, (2022).

45. Gatti, G. *et al.* Flat Γ Moiré Bands in Twisted Bilayer WSe2. *Phys Rev Lett* **131**, (2023).

46. Naik, M. H. & Jain, M. Origin of layer dependence in band structures of two-dimensional materials. *Phys Rev B* **95**, (2017).

47. Li, H. *et al.* Imaging moiré flat bands in three-dimensional reconstructed WSe2/WS2 superlattices. *Nat Mater* **20**, 945–950 (2021).

48. Kundu, S., Naik, M. H., Krishnamurthy, H. R. & Jain, M. Moiré induced topology and flat bands in twisted bilayer WSe2: A first-principles study. *Phys Rev B* **105**, (2022).

49. Naik, M. H. *et al.* Intralayer charge-transfer moiré excitons in van der Waals superlattices. *Nature* **609**, 52–57 (2022).

50. Zheng, H. *et al.* Strong Interlayer Coupling in Twisted Transition Metal Dichalcogenide Moiré Superlattices. *Advanced Materials* **35**, (2023).

51. Kennes, D. M. *et al.* Moiré heterostructures as a condensed-matter quantum simulator. *Nature Physics* vol. 17 155–163 Preprint at https://doi.org/10.1038/s41567-020-01154-3 (2021).

52. Graham, A. J. *et al.* Conduction Band Replicas in a 2D Moiré Semiconductor Heterobilayer. *Nano Lett* **24**, 5117–5124 (2024).

53. Polley, C. M. *et al.* Origin of the π -band replicas in the electronic structure of graphene grown on 4H -SiC(0001). *Phys Rev B* **99**, (2019).

54. Amorim, B. General theoretical description of angle-resolved photoemission spectroscopy of van der Waals structures. *Phys Rev B* **97**, (2018).

55. Rostami, H. *et al.* Layer and orbital interference effects in photoemission from transition metal dichalcogenides. *Phys Rev B* **100**, (2019).

56. Ulstrup, S. *et al. Direct Observation of Minibands in a Twisted Graphene/WS 2 Bilayer*. https://www.science.org (2020).

57. Razzoli, E. *et al.* Selective Probing of Hidden Spin-Polarized States in Inversion-Symmetric Bulk MoS2. *Phys Rev Lett* **118**, (2017).

58. Schüler, M. *et al.* Polarization-Modulated Angle-Resolved Photoemission Spectroscopy: Toward Circular Dichroism without Circular Photons and Bloch Wave-function Reconstruction. *Phys Rev X* **12**, (2022).

59. Ghiotto, A. *et al.* Quantum criticality in twisted transition metal dichalcogenides. *Nature* **597**, 345–349 (2021).